\documentclass[letterpaper, 10 pt, conference]{ieeeconf}
\pdfoutput=1

\IEEEoverridecommandlockouts
\usepackage{amsmath,amssymb}
\usepackage{iftex}
\ifXeTeX
  \usepackage{fontspec}
\fi

\usepackage{amsthm}
\usepackage{capt-of}
\usepackage{microtype}
\usepackage{graphicx}
\graphicspath{{./}}
\usepackage{booktabs}
\usepackage{array}
\usepackage{multirow}
\usepackage{xcolor}
\usepackage{algorithm}
\usepackage{algpseudocode}
\floatname{algorithm}{Algorithm}
\algrenewcommand\algorithmicrequire{\textbf{Input:}}
\algrenewcommand\algorithmicensure{\textbf{Output:}}
\newcommand{\BGStage}[1]{\Statex\textbf{#1}}
\usepackage{url}
\usepackage{tikz}
\usetikzlibrary{arrows.meta,positioning}
\usepackage{pgfplots}
\pgfplotsset{compat=1.18}
\usepackage[hidelinks]{hyperref}

\theoremstyle{definition}
\newtheorem{definition}{Definition}
\newtheorem{example}{Example}

\newcommand{\method}{FlatSeek}

\fussy
\makeatletter
\renewcommand\paragraph{\@startsection{paragraph}{4}{0pt}%
  {-3.25ex\@plus -1ex \@minus -.2ex}%
  {1.5ex \@plus .2ex}%
  {\normalfont\bfseries}}
\makeatother

\title{\LARGE\bf
FlatSeek: LLM-Driven Symbolic Discovery of Flat Outputs via Backward Data Generation}

\author{Zihao Fu$^{*}$, Xingchen Li$^{*}$, Keyou You%
\thanks{$^{*}$These authors contributed equally to this work (Corresponding author: Keyou You). This work was supported by National Natural Science Foundation of China (62325305).}%
\thanks{The authors are with the Department of Automation, and Beijing National Research Center for Info. Sci. \& Tech. (BNRist), Tsinghua University, Beijing 100084, China. (e-mail: \{fuzh24,lixc21\}@mails.tsinghua.edu.cn, youky@tsinghua.edu.cn)}%
}

\begin{document}
\maketitle
\thispagestyle{empty}
\pagestyle{empty}

\begin{abstract}
Differential flatness enables efficient planning and control, but discovering explicit flat outputs for general nonlinear systems remains difficult due to the lack of general methods and supervised data. We propose FlatSeek, a framework combining backward data generation with task-specific large language model (LLM) training. By reversing the defining reconstruction of flatness, we generate system--flat-output pairs with known certificates, then fine-tune an LLM to propose symbolic flat-output candidates for unseen dynamics.
On literature and synthetic benchmarks, \method{} outperforms both the traditional symbolic methods and the untuned reference models in flat-output discovery success, and uncovers alternative symbolic flat outputs not reported in the literature.
\end{abstract}

\section{Introduction}\label{sec:intro}

A control system is differentially flat if it admits an output, called a \emph{flat output}, from which all states and inputs can be reconstructed through smooth maps using finitely many output derivatives~\cite{fliess1995flatness,levine2009flatness}.
Once a flat output is known, a sufficiently smooth output curve satisfying prescribed boundary conditions can be mapped to feasible state and input trajectories by differentiation and evaluation of the reconstruction maps, without integrating the system dynamics~\cite{martin2003flatness}.
Differential flatness has thus become an important framework for planning and control, with applications to quadrotors~\cite{mellinger2011minimum} as well as mechanical and power electronic systems~\cite{sira2004flat}.

Finding a flat output, however, remains difficult: no general procedure maps arbitrary dynamics to one~\cite{levine2009flatness}. Flat outputs have no fixed form and, for some systems, necessarily involve input derivatives of an unknown order~\cite{martin2003flatness,gstottner2023noxu}. The suitable form must therefore be found system by system, and manual trial and error is not a reliable general procedure.

Existing approaches construct flat outputs under static or dynamic feedback linearization, dynamic extension, or pure input prolongation~\cite{hunt1983global,charlet1989dynamic,levine2025prolongation}, and each applies only when its corresponding structural conditions hold. Complementary studies derive necessary conditions, differential-geometric tests, or order bounds~\cite{pomet2009necessary,kaminski2024jacobi}, which can screen candidates without constructing outputs. However, they offer little guidance when the form of the flat output is unknown and the required structural conditions do not hold.

Learning-based methods have also been applied to flat-output identification. For example, neural networks have been used to identify flat outputs of linearized underactuated mechanical systems from measured data~\cite{ma2024flatnn}. However, an output identified for the linearization holds only locally and need not be flat for the original nonlinear system. Symbolic flat-output discovery for nonlinear systems therefore remains an open challenge.

Large language models (LLMs) provide a natural starting point for this search. Pretrained on large amounts of mathematical text and code, they can already read differential equations and write symbolic expressions in the variables involved. Prior work also shows that they can generate explicit symbolic and mathematical structures~\cite{alfarano2024lyapunov,deepmind2024funsearch,zhong2026firstintegrals}.

We propose \method{}, a framework that fine-tunes an LLM to propose explicit symbolic flat-output candidates for unseen dynamics and screens them with a verifier. Fine-tuning requires system--flat-output pairs, but, to our knowledge, no public corpus provides them. Labeling sampled systems does not help either, since labeling is itself the discovery problem we aim to solve. The reverse direction, from output to dynamics, is constructive: once a flat output and its reconstruction maps are fixed, compatible dynamics follow by differential closure, with the output known by construction. \method{} therefore generates the pairs backward, diversifying them with invertible coordinate transformations that preserve the known flat parametrizations, and fine-tunes the LLM on them.

Our contributions are threefold. First, we turn the reconstruction structure of differential flatness into a controllable source of paired training examples through backward dataset construction.
Second, we propose \method{}, a framework that integrates backward data generation, task-specific language-model training, and inference-time candidate search for symbolic flat-output discovery from unseen dynamics.
Third, we evaluate flat-output discovery success rates on literature and synthetic benchmarks against traditional symbolic methods and untuned reference models, and discover alternative symbolic flat outputs not reported in the literature.

Section~\ref{sec:preliminaries} formulates the problem, Sections~\ref{sec:bg-generation} and~\ref{sec:train} present the backward construction and the LLM training and search, Section~\ref{sec:exp} reports the evaluation, and Section~\ref{sec:conclusion} concludes the paper.

  \section{\texorpdfstring{Problem Formulation and Preliminaries}{Problem Formulation and Preliminaries}}

To ground our approach, this section reviews differential flatness and frames explicit flat-output discovery as a conditional symbolic generation task.

\label{sec:preliminaries}

\subsection{Differential Flatness}
\label{sec:pf-notation}

Consider a nonlinear control system
\begin{equation}
    \dot{x}=F(x,u),
    \qquad
    x\in\mathbb{R}^{n},\quad
    u\in\mathbb{R}^{m},
    \label{eq:pf-system}
\end{equation}
where $F:\mathcal X\times\mathcal U\to\mathbb{R}^{n}$ is $C^\infty$ on
the open domain $\mathcal X\times\mathcal U$, with
$\mathcal X\subset\mathbb{R}^{n}$ and $\mathcal U\subset\mathbb{R}^{m}$.

\begin{definition}[Differential flatness
{\cite{levine2009flatness,sira2004flat}}]
\label{def:flatness}
System~\eqref{eq:pf-system} is differentially flat if there exist
finite nonnegative integers $r$ and $q$ and a smooth $m$-dimensional output
map
\begin{equation}
    y=h\!\left(x,u,\dot{u},\ldots,u^{(r)}\right),
    \qquad
    y\in\mathbb{R}^{m},
    \label{eq:pf-output}
\end{equation}
whose components are differentially independent, and smooth
reconstruction maps $\phi$ and $\psi$ exist such that all states and inputs
can be expressed in terms of $y$ and finitely many of its time derivatives:
\begin{equation}
\begin{aligned}
    x &= \phi\!\left(y,\dot{y},\ldots,y^{(q)}\right),\\
    u &= \psi\!\left(y,\dot{y},\ldots,y^{(q)}\right).
\end{aligned}
\label{eq:pf-reconstruction}
\end{equation}
Conversely, every sufficiently smooth curve $y(t)$ for which these
expressions are defined, when substituted into~\eqref{eq:pf-reconstruction},
yields a trajectory satisfying~\eqref{eq:pf-system}.
The output $y$ is then called a \emph{flat output}.
\end{definition}
The common order $q$ is an upper bound; the two reconstruction maps may use different derivative orders.

\begin{center}
\begin{minipage}{\columnwidth}
    \centering
    \includegraphics[width=\columnwidth,trim=0bp 88bp 0bp 84bp,clip]{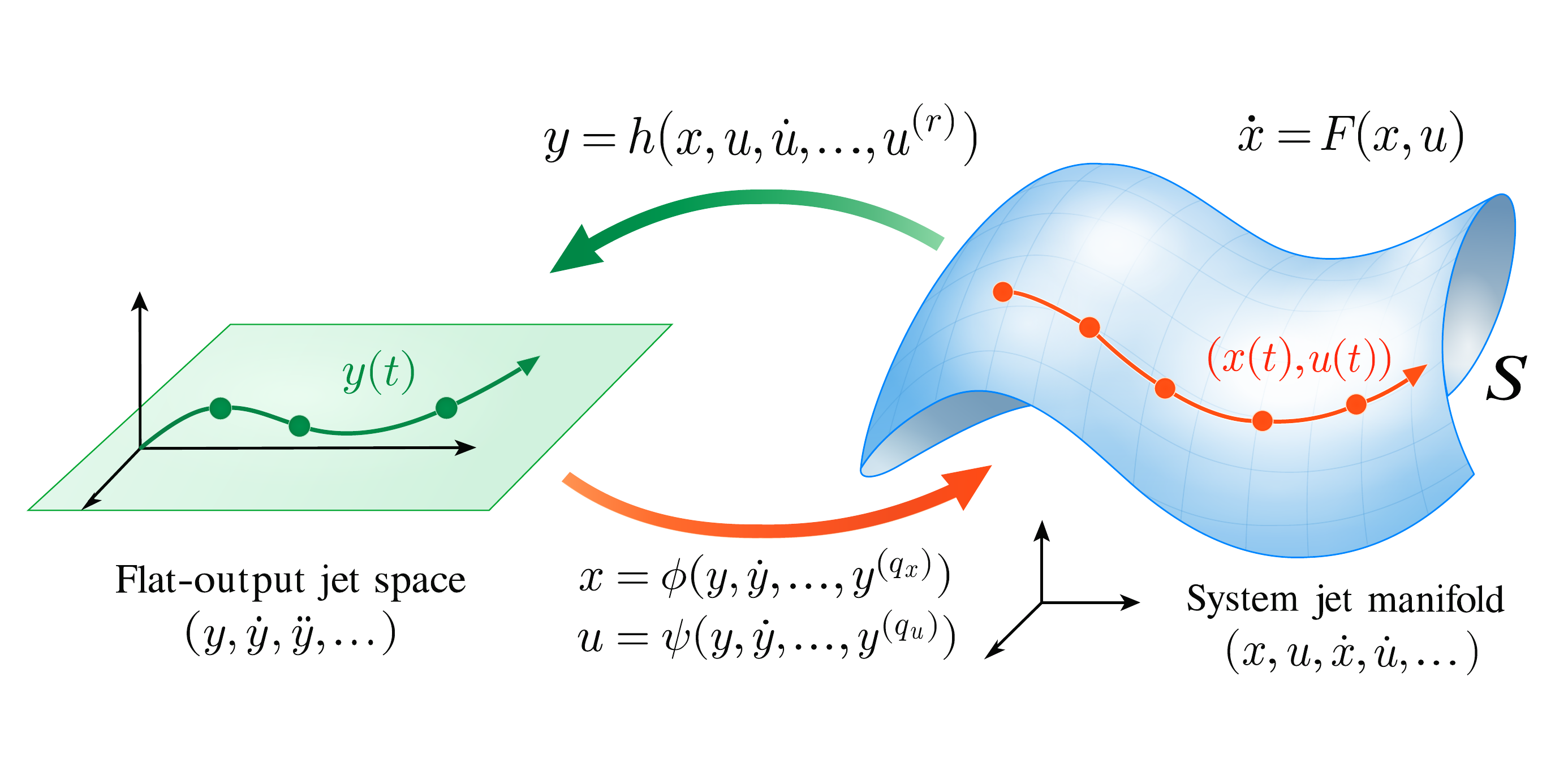}
    \captionof{figure}{Schematic of differential flatness. In a regular local region where the maps are valid, flat-output derivatives reconstruct states and inputs. Curves show trajectory jet lifts, i.e., trajectories together with their time derivatives; arrows show output/reconstruction maps and temporal prolongation.}
    \label{fig:differential-flatness}
\end{minipage}
\end{center}

Differential flatness therefore provides a finite differential
parametrization of system trajectories, as illustrated in
Fig.~\ref{fig:differential-flatness}: once a flat-output trajectory
$y(t)$ is specified, the corresponding states and inputs follow directly
from $y$ and its derivatives through~\eqref{eq:pf-reconstruction}, without
integrating the original dynamics.

\subsection{\texorpdfstring{Flat-Output Discovery as Symbolic Generation}{Flat-Output Discovery as Symbolic Generation}}
\label{sec:llm-symbolic}

This parametrization presupposes a known flat output; for general nonlinear systems, finding one is the difficult step. Given the dynamics $F$, we seek an explicit symbolic output map $h$ for which Definition~\ref{def:flatness} holds, and formulate this as conditional symbolic generation: the model receives a symbolic description of $F$ and proposes a candidate flat output $h$. Both are represented as token sequences by
$\sigma(\cdot)=\operatorname{Tok}\circ\operatorname{Ser}(\cdot)$, where
$\operatorname{Ser}$ serializes an expression into deterministic text and
$\operatorname{Tok}$ tokenizes it.

For $\sigma(h)=(\tau_1,\ldots,\tau_{L_h})$, where $L_h$ denotes the sequence
length and $\tau_t$ is the $t$-th token, an autoregressive language model
defines
\begin{equation}
\pi_\theta\!\left(
    \sigma(h)\mid\sigma(F)
\right)
=
\prod_{t=1}^{L_h}
\pi_\theta\!\left(
    \tau_t\mid\sigma(F),\tau_{<t}
\right),
\label{eq:llm-autoregressive}
\end{equation}
where $\tau_{<t}=(\tau_1,\ldots,\tau_{t-1})$. Sampling this distribution
yields multiple candidate outputs for the same dynamics, and a
symbolic verifier checks whether each candidate satisfies the required
flatness conditions.

We fine-tune the model for this task using supervised system--output pairs
$\mathcal D=\{(F^{(i)},h^{(i)})\}_{i=1}^{N}$. Constructing such supervision
is challenging because a valid label $h^{(i)}$ must already be known for each
sampled dynamics $F^{(i)}$. The next section addresses this difficulty by
constructing the pairs in reverse: it starts from finite-derivative
reconstruction maps and derives compatible dynamics with known flat outputs.
\section{Backward Construction of Flat Systems}
\label{sec:bg-generation}
\label{sec:bg-principle}

\begin{figure*}[t]
    \vspace{-6pt}
    \centering
    \includegraphics[width=\textwidth,trim=15bp 57bp 9.5bp 112.2bp,clip]{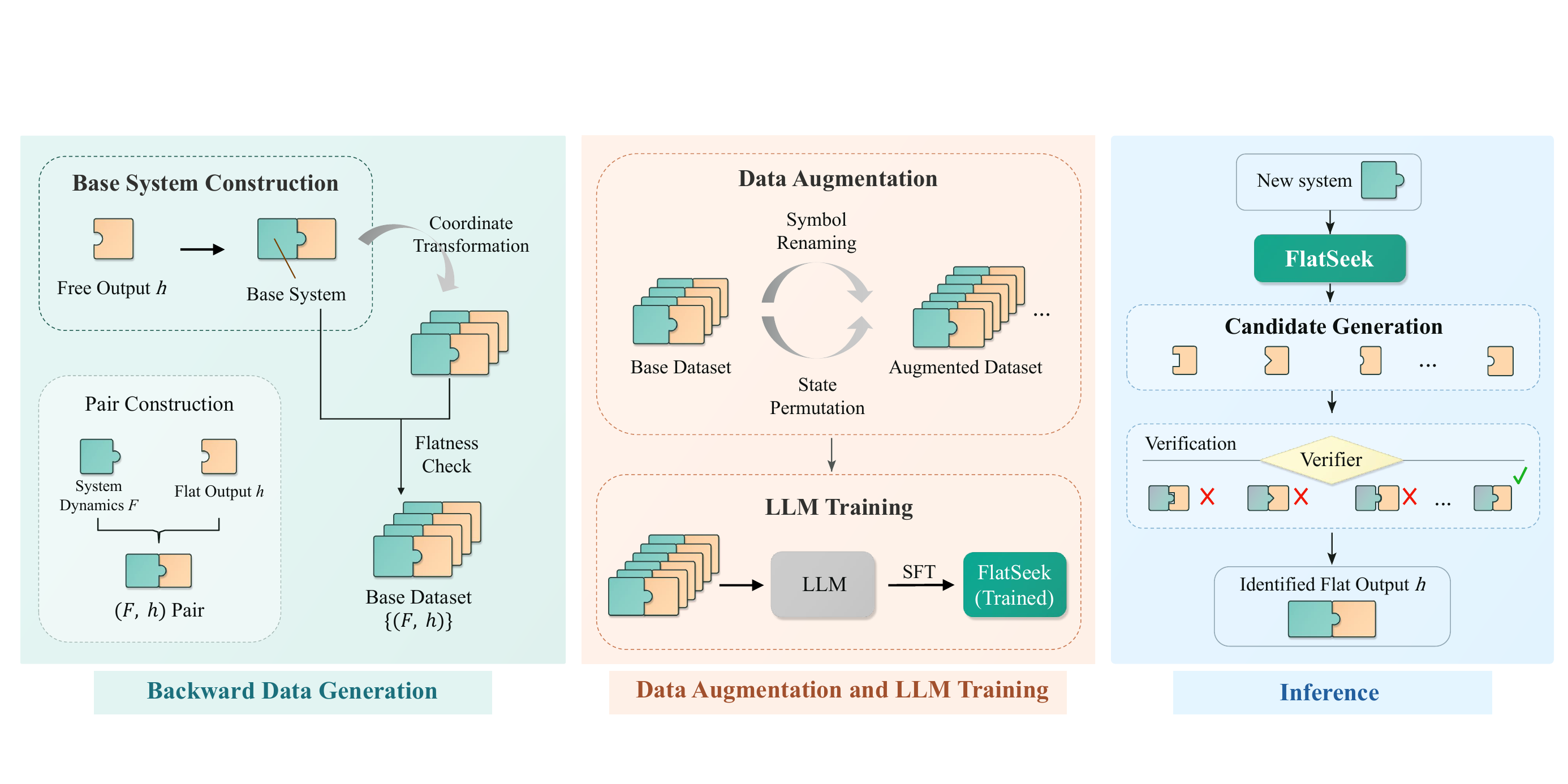}
    \caption{Overview of \method{}'s three stages: (1) backward data generation of system--flat-output pairs $(F,h)$, (2) data augmentation and LLM training, and (3) inference, in which the trained model proposes candidates that the verifier screens.}
    \label{fig:method-overview}
    \vspace{-6pt}
\end{figure*}

As shown in stage~1 of Fig.~\ref{fig:method-overview},
this section constructs the pairs $\mathcal D$ in two parts: base systems
whose flat parametrizations are known by design, and invertible coordinate
transformations that diversify their symbolic forms while preserving these
parametrizations.

\subsection{\texorpdfstring{Base Systems from Output Parametrizations}{Base Systems from Output Parametrizations}}
\label{sec:bg-family}

A base system is obtained by prescribing how a free output $y$ and
finitely many of its derivatives determine the state and input, and then
deriving compatible dynamics $\dot z=F_0(z,u)$. Since this construction
repeatedly differentiates functions of $y$ and its derivatives, we write the
total time derivative for $Y_k=(y,\dot y,\ldots,y^{(k)})$ as
\[
D_y g
=
\sum_{j=0}^{k}
\frac{\partial g}{\partial y^{(j)}}y^{(j+1)},
\qquad
g=g(Y_k).
\]

\emph{Step 1: Construct the output parametrization.}
We first specify how the output and its derivatives will
reconstruct the state and input of the system to be generated.
Choose $y=(y_1,\ldots,y_m)^\top$ and chain lengths $\ell_i\ge1$.
Each chain assigns successive output derivatives to state coordinates and its
terminal derivative to the corresponding input. Let
\[
\begin{gathered}
n=\sum_{i=1}^{m}\ell_i,\qquad
q=\max_{1\le i\le m}\ell_i,\\[2pt]
\mathcal I=\left\{(i,k):1\le i\le m,\ 0\le k<\ell_i\right\}.
\end{gathered}
\]
Analytic weights $a_{i,k}(y,\dot y_1)$, $(i,k)\in\mathcal I$, nonzero on
the selected regular domain, introduce the nonlinear couplings of the
generated dynamics. The first chain is left unweighted so that $\dot y_1$ is
itself a state or input coordinate, and terminal weights are set to one,
since they could be absorbed into the inputs:
\[
 a_{1,k}\equiv1\quad(0\le k<\ell_1),
 \qquad
 a_{i,\ell_i-1}\equiv1\quad(1\le i\le m).
\]
Define the weighted derivative chains
\begin{equation}
\zeta_{i,0}=y_i,
\qquad
\zeta_{i,k+1}
=
\frac{D_y\zeta_{i,k}}
     {a_{i,k}(y,\dot y_1)},
\qquad
0\le k<\ell_i.
\label{eq:bg-chain}
\end{equation}
The first $\ell_i$ elements of each chain are assigned to the state, while
the terminal element is assigned to the corresponding input:
\begin{equation}
\begin{aligned}
z=&\phi_0(Y_{q-1})
=
\operatorname{col}_{i=1}^{m}
(\zeta_{i,0},\ldots,\zeta_{i,\ell_i-1}),\\
u=&\psi_0(Y_q)
=
(\zeta_{1,\ell_1},\ldots,\zeta_{m,\ell_m})^\top .
\end{aligned}
\label{eq:bg-seed}
\end{equation}
Here $\operatorname{col}$ stacks the listed chain entries into a column
vector.
Writing these coordinates as $z_{i,k}$ and setting
$z_{i,\ell_i}:=u_i$, the chain heads define
\begin{equation}
h_0(z)
=
(z_{1,0},\ldots,z_{m,0})^\top .
\label{eq:bg-base-label-general}
\end{equation}

\emph{Step 2: Derive the compatible dynamics.}
We next derive a state equation that the prescribed state and
input trajectories satisfy.
Since the first chain is unweighted,
\[
b(z,u):=z_{1,1},
\qquad
b(\phi_0,\psi_0)=\dot y_1.
\]
Differential closure replaces the derivative of each chain state by the next
chain element. Substituting $y=h_0(z)$ and $\dot y_1=b(z,u)$ into~\eqref{eq:bg-chain}
therefore yields
\begin{equation}
\dot z_{i,k}
=
a_{i,k}\!\left(h_0(z),b(z,u)\right)z_{i,k+1},
  \qquad
  (i,k)\in\mathcal I,
\label{eq:bg-explicit-system}
\end{equation}
which defines $\dot z=F_0(z,u)$. By construction,
\begin{equation}
D_y\phi_0=F_0(\phi_0,\psi_0),
\qquad
h_0(\phi_0)=y.
\label{eq:bg-principle}
\end{equation}

\emph{Step 3: Recover the parametrization.}
We finally verify that the chosen output allows all states and
inputs of the generated system to be recovered from finitely many output
derivatives.
On the regular domain
\[
\Omega_0
=
\left\{
(z,u):
  a_{i,k}\!\left(h_0(z),b(z,u)\right)\ne0,
  \quad \forall\,(i,k)\in\mathcal I
\right\},
\]
the recursive inversion is
\begin{equation}
 z_{i,k+1}
 =
 \frac{\dot z_{i,k}}
      {a_{i,k}(y,\dot y_1)},
 \qquad (i,k)\in\mathcal I.
 \label{eq:bg-recovery-recursion}
\end{equation}
Thus, starting from $z_{i,0}=y_i$, all states and inputs are recovered from
finitely many derivatives of $y$, reproducing $\phi_0(Y_{q-1})$ and
$\psi_0(Y_q)$.
Although $\partial\phi_0/\partial Y_{q-1}$ need not be square, its $n\times n$ submatrix corresponding to the chain coordinates is triangular after ordering them by derivative order, with nonzero diagonal entries $\prod_{s=0}^{k-1}a_{i,s}^{-1}$; hence it has rank $n$ on $\Omega_0$. Together with~\eqref{eq:bg-principle}, this establishes the required local reconstruction and shows that $h_0$ is a flat output of $F_0$.

\begin{example}[Backward construction of a four-dimensional base system]
\label{sec:bg-base-example}

Take $m=2$, $(\ell_1,\ell_2)=(1,3)$, and
\[
a_{2,0}=a_{2,1}=\dot y_1,
\qquad
a_{2,2}=1.
\]

For $\dot y_1\ne0$, let $\eta=\dot y_2/\dot y_1$ and order the states as
\[
(z_1,z_2,z_3,z_4)
=
(z_{1,0},z_{2,2},z_{2,1},z_{2,0}).
\]
Then~\eqref{eq:bg-chain} gives
\begin{equation}
\begin{aligned}
\phi_0(Y_2)
&=
\left(
y_1,\,
\frac{D_y\eta}{\dot y_1},\,
\eta,\,
y_2
\right)^\top,\\
\psi_0(Y_3)
&=
\left(
\dot y_1,\,
D_y\!\left(
\frac{D_y\eta}{\dot y_1}
\right)
\right)^\top .
\end{aligned}
\label{eq:bg-chain-def}
\end{equation}

Equation~\eqref{eq:bg-explicit-system} becomes
\begin{equation}
\begin{aligned}
\dot z_1&=u_1,
&
\dot z_2&=u_2,\\
\dot z_3&=z_2u_1,
&
\dot z_4&=z_3u_1.
\end{aligned}
\label{eq:bg-chain-four}
\end{equation}
The corresponding base flat output is $h_0(z)=(z_1,z_4)^\top$. Thus
$h_0$ is a flat output on $\dot y_1\ne0$, with state and input reconstruction
given by $\phi_0(Y_2)$ and $\psi_0(Y_3)$, respectively.
\end{example}

\subsection{\texorpdfstring{Symbolic Diversification via Coordinate Transformations}{Symbolic Diversification via Coordinate Transformations}}
\label{sec:bg-transport}

The base construction guarantees flatness but produces expressions closely
tied to the chain coordinates. We therefore apply invertible state
transformations only to diversify the symbolic representations while
updating the known reconstruction maps consistently.

Let $x=T(z)$ be an analytic diffeomorphism with an explicit inverse and a Jacobian $J_T(z):=\partial T/\partial z$.
The transformed dynamics, output, and reconstruction maps are
\begin{equation}
\begin{aligned}
F(x,u)
&=
J_T\!\left(T^{-1}(x)\right)
F_0\!\left(T^{-1}(x),u\right),\\
h(x)
&=
h_0\!\left(T^{-1}(x)\right),\\
\phi
&=
T\circ\phi_0,
\qquad
\psi=\psi_0.
\end{aligned}
\label{eq:bg-transport}
\end{equation}
Since $T$ is invertible, the base parametrization is inherited directly:
\begin{equation}
D_y\phi=F(\phi,\psi),\qquad h(\phi)=y.
\label{eq:bg-transport-identities}
\end{equation}
Hence $h$ remains a flat output of the transformed system. Its regular
domain is $\Omega=\{(x,u):(T^{-1}(x),u)\in\Omega_0\}$.
The resulting labels $h(x)$ depend on the state only.

Since~\eqref{eq:bg-transport} uses $T^{-1}$ explicitly, a convenient
choice is the shear
\begin{equation}
\begin{aligned}
T_S(z)&=z+e_jS(z_{-j}),\\
T_S^{-1}(x)&=x-e_jS(x_{-j}),
\end{aligned}
\label{eq:bg-shear}
\end{equation}
 where $e_j$ is the $j$th standard basis vector, $z_{-j}$ denotes the
 vector obtained by removing the $j$th component of $z$, and $S$ is a
 scalar analytic function of those $n-1$ coordinates. Thus $S$ does not
 depend on the modified coordinate $z_j$.
Multiple shears may be composed to further increase symbolic diversity.

\begin{example}[Coordinate transformation of the four-dimensional base system]
\label{sec:bg-vehicle-example}

Continuing Example~\ref{sec:bg-base-example}, choose
\[
j=4,
\qquad
S(z_{-4})=z_1z_3^2.
\]
Then
\[
x_4=z_4+z_1z_3^2,
\]
with the remaining coordinates unchanged. Applying~\eqref{eq:bg-transport}
to~\eqref{eq:bg-chain-four} gives
\begin{equation}
\begin{aligned}
\dot x_1&=u_1,
&
\dot x_2&=u_2,\\
\dot x_3&=x_2u_1,
&
\dot x_4&=
x_3(1+x_3+2x_1x_2)u_1,
\end{aligned}
\label{eq:bg-example}
\end{equation}
while the flat output becomes
\begin{equation}
h(x)
=
\left(
x_1,\,
x_4-x_1x_3^2
\right)^\top.
\label{eq:bg-label}
\end{equation}
The coordinate change thus turns the simple base output $(z_1,z_4)^\top$
into a nonlinear expression with the same flat parametrization.
\end{example}

Algorithm~\ref{alg:bg-dataset} summarizes the complete backward-generation
pipeline. Each accepted configuration produces a supervised pair $(F,h)$
together with the reconstruction maps $(\phi,\psi)$, their regular domain
$\Omega$, and derivative-order information $q$.

\begin{algorithm}[t]
\caption{Backward data generation}
\label{alg:bg-dataset}
\small
\begin{algorithmic}[1]

\Require Configuration set $\mathcal G$ specifying
$m$, $(\ell_i)$, $\{a_{i,k}\}$, and $T$
\Ensure Supervision set $\mathcal D$ with reconstruction maps

\State $\mathcal D\gets\varnothing$

\For{each $\gamma\in\mathcal G$}

    \BGStage{Recursive construction}

    \State Set $\zeta_{i,0}=y_i$ and recursively compute
    \Statex \hspace{\algorithmicindent}$\displaystyle
    \zeta_{i,k+1}
    =
    \frac{D_y\zeta_{i,k}}
         {a_{i,k}(y,\dot y_1)},
    \qquad
    0\le k<\ell_i$

    \State Form $\phi_0$ from the first $\ell_i$ entries of each
    chain and $\psi_0$ from the terminal entries; set $q=\max_i\ell_i$

    \BGStage{Dynamic closure}

    \State Set $h_0(z)=(z_{1,0},\ldots,z_{m,0})^\top$
    \State With $z_{i,\ell_i}:=u_i$ and $b(z,u):=z_{1,1}$, define
    \Statex \hspace{\algorithmicindent}$\displaystyle
    (F_0)_{i,k}
    =
    a_{i,k}\!\left(h_0(z),b(z,u)\right)z_{i,k+1}$
    \State Define $\Omega_0$ where all recursive weights are nonzero

    \BGStage{Coordinate transformations}

    \State Apply the coordinate transformation $x=T(z)$
    \Statex \hspace{\algorithmicindent}$\displaystyle
    F(x,u)
    =
    J_T\!\left(T^{-1}(x)\right)
    F_0\!\left(T^{-1}(x),u\right)$
    \State Set $h=h_0\circ T^{-1}$, $\phi=T\circ\phi_0$,
    and $\psi=\psi_0$
    \State Map $\Omega_0$ under $(z,u)\mapsto(T(z),u)$
    to obtain $\Omega$

    \BGStage{Flatness check}

    \State Check the reconstruction identities and rank condition on $\Omega$

    \If{$\Omega\ne\varnothing$ and the conditions are verified}
        \State Add $(F,h)$ to $\mathcal D$ and store $(\phi,\psi,\Omega,q)$
    \EndIf

\EndFor

\State Canonicalize the accepted $(F,h)$ pairs and remove duplicates
\State \Return $\mathcal D$ and the corresponding reconstruction maps

\end{algorithmic}
\end{algorithm}

\section{\texorpdfstring{LLM Training and Search for Flat Outputs}{LLM Training and Search for Flat Outputs}}
\label{sec:train}

Building on the pairs constructed in
Section~\ref{sec:bg-generation}, \method{} trains and deploys the LLM in two
stages. Augmenting the pairs first enlarges the training data at no
labeling cost, so that fine-tuning teaches the model to read the dynamics
independently of variable names and state order (stage~2 in
Fig.~\ref{fig:method-overview}). At inference, the tuned model proposes
multiple candidates and a symbolic verifier screens them, so the model can
explore diverse flat outputs while every accepted one passes an explicit
check (stage~3).

\subsection{\texorpdfstring{Data Augmentation and LLM Fine-Tuning}{Data Augmentation and LLM Fine-Tuning}}
\label{sec:train-supervision}

To let the model learn the dynamics independently of notation, we
augment the pairs in two ways. First, let $\rho$ be a bijection on the declared state and
input symbols, applied consistently to the system description, derivative
symbols, and output expression:
\begin{equation}
(F,h)
\longmapsto
\bigl(\rho(F),\rho(h)\bigr).
\label{eq:rename-augmentation}
\end{equation}

Second, let $P\in\{0,1\}^{n\times n}$ be a state-permutation matrix with
$P^{-1}=P^\top$. Reordering the states as $\widetilde x=Px$ gives
\begin{equation}
\begin{aligned}
\widetilde F(\widetilde x,u)
&=P F(P^\top\widetilde x,u),\\[-1pt]
\widetilde h(\widetilde x)
&=h(P^\top\widetilde x).
\end{aligned}
\label{eq:permutation-augmentation}
\end{equation}
Neither operation changes the underlying system, so each label
remains a valid flat output without being recomputed. Applied separately or
jointly, they enlarge the training set at no additional labeling cost.

Using the resulting pairs, we learn the conditional mapping
$\sigma(F)\mapsto\sigma(h)$. For a minibatch $\mathcal B$ and a target
sequence $\sigma(h)=(\tau_1,\ldots,\tau_{L_h})$, we minimize the
supervised fine-tuning (SFT) loss, i.e., the answer-token negative log-likelihood:
\begin{equation}
\mathcal L_{\mathrm{SFT}}(\theta)
=
-
\frac{
\displaystyle
\sum_{(F,h)\in\mathcal B}
\sum_{t=1}^{L_h}
\log\pi_\theta\!\left(
\tau_t\mid\sigma(F),\tau_{<t}
\right)
}{
\displaystyle
\sum_{(F,h)\in\mathcal B}L_h
}.
\label{eq:sft}
\end{equation}
Prompt tokens are masked from the loss. We use LoRA~\cite{hu2021lora},
freezing the base-model weights and updating only low-rank adapters,
so that the pretrained mathematical knowledge is retained.

\subsection{\texorpdfstring{Candidate Generation and Verification}{Candidate Generation and Verification}}
\label{sec:train-symbolic}

For an unseen $m$-input system $F$, the trained model takes its symbolic
description $\sigma(F)$ as input and generates flat-output candidates
autoregressively.
Since flat outputs are not unique and any verified one suffices,
we allocate a budget of $K$ candidates to each system and sample an
ordered sequence:
\begin{equation}
\sigma\!\left(\hat h^{(k)}\right)
\sim\pi_\theta\!\left(\cdot\mid\sigma(F)\right),
\qquad k=1,\ldots,K.
\label{eq:candidate-set}
\end{equation}
Each generated sequence is parsed into an $m$-dimensional symbolic
expression. The model thus only needs to propose plausible
candidates, while acceptance is decided solely by the verifier, which
differentiates each candidate along the dynamics and applies a finite-order Jacobian-rank test for recoverability of the state and input from the output derivatives; candidates with differentially dependent components are rejected.

\begin{figure}[t]
\centering
\begin{tikzpicture}[
  panel/.style={
    draw=black!22,
    fill=black!4,
    rounded corners=2pt,
    line width=0.55pt,
    inner xsep=8pt,
    inner ysep=7pt,
    text width=\columnwidth-34pt,
    align=left,
    text=black!95,
    font=\ttfamily\fontsize{7.5}{9}\selectfont
  },
  arrow/.style={
    -{Stealth[length=2mm,width=1.4mm]},
    draw=black!55,
    line width=0.85pt,
    shorten <=0.8pt,
    shorten >=0.8pt
  },
  arrowlabel/.style={
    font=\ttfamily\fontsize{7.5}{9}\selectfont,
    text=black!55,
    fill=none,
    inner sep=0pt,
    right
  }
]

\node[panel] (prompt) {\raggedright
  {\bfseries Prompt}\par\vspace{3pt}
  {\color{black!18}\hrule height 0.5pt}\vspace{5pt}

  {\mdseries SYSTEM}\\[1pt]
  Given nonlinear dynamics $F(x,u)$, propose an $m$-dimensional flat output.
  Return exactly $m$ symbolic expressions. Input derivatives, when needed,
  may be declared using ujet.\par

  \vspace{5pt}
  {\mdseries USER}\\[1pt]
  States: x1, x2, x3, x4\par
  Inputs: u1, u2\par
  \vspace{1pt}
  dx1/dt = u1\par
  dx2/dt = u2\par
  dx3/dt = x2*u1\par
  dx4/dt = x3*(1+x3+2*x1*x2)*u1\par
  \vspace{1pt}
  Find a flat output of dimension m=2.
};

\node[
  panel,
  below=7mm of prompt.south,
  anchor=north
] (out) {\raggedright
  {\bfseries Model output}\par\vspace{3pt}
  {\color{black!18}\hrule height 0.5pt}\vspace{5pt}
  \{x4-x1*x3**2,x1\}
};

\draw[arrow] (prompt.south) -- (out.north)
  node[arrowlabel, midway, xshift=6pt]
  {sample K candidates};

\end{tikzpicture}

\caption{Prompt interface and one generated candidate. The dynamics are from
Example~\ref{sec:bg-vehicle-example}; the candidate equals $h$
in~\eqref{eq:bg-label} up to component order.}
\label{fig:prompt}
\end{figure}

\section{Evaluation on Literature and Synthetic Benchmarks}\label{sec:exp}

We evaluate flat-output discovery on the literature and synthetic benchmarks. All methods use the candidate-checking protocol described in Section~\ref{sec:train-symbolic}, and a system is counted as solved when at least one candidate is accepted. We compare five traditional symbolic methods, four untuned open-source models, and an API reasoning model, and report a candidate-budget ablation.

\begin{table}[t]
\caption{Systems with an accepted candidate on each benchmark}
\label{tab:main}
\centering\footnotesize
\setlength{\tabcolsep}{2pt}
\renewcommand{\arraystretch}{1.0}
\begin{tabular}{@{}>{\raggedright\arraybackslash}m{0.19\columnwidth}
                >{\raggedright\arraybackslash}m{0.38\columnwidth}
                >{\centering\arraybackslash}m{0.15\columnwidth}
                >{\centering\arraybackslash}m{0.15\columnwidth}@{}}
\toprule
\textbf{Category} & \textbf{Method} &
\shortstack{\textbf{Literature}\\[-1pt]\scriptsize$(N=83)$} &
\shortstack{\textbf{Synthetic}\\[-1pt]\scriptsize$(N=200)$}\\
\midrule
\multirow{5}{*}{\shortstack[l]{\scriptsize\emph{Traditional}\\[-1pt]\scriptsize\emph{symbolic}\\[-1pt]\scriptsize\emph{methods}}}
  & Moving-frame construction & 8 & 2\\
  & Static feedback linearization & 45 & 45\\
  & Saddle--Jacobi screening & 47 & 120\\
  & Bounded pure prolongation & 54 & 127\\
  & State-coordinate enumeration & 58 & 112\\
\midrule
\multirow{5}{*}{\shortstack[l]{\scriptsize\emph{Untuned}\\[-1pt]\scriptsize\emph{reference}\\[-1pt]\scriptsize\emph{models}}}
  & Qwen3.5-2B & 50 & 28\\
  & Qwen3.5-9B & 59 & 88\\
  & Qwen3.5-27B & 61 & 89\\
  & R1-Distill-32B & 54 & 66\\
  & DeepSeek V4 Flash (API) & 67 & 97\\
\midrule
\scriptsize\emph{Proposed method}
  & \textbf{\method{}} & \textbf{74} & \textbf{171}\\
\bottomrule
\end{tabular}
\end{table}

\begin{figure}[t]
\centering
\begin{tikzpicture}[
  card/.style={
    draw=black!18,
    line width=0.4pt,
    rounded corners=2pt,
    fill=black!3,
    inner xsep=6pt,
    inner ysep=5pt,
    text width=\columnwidth-14pt,
    align=left
  }
]
\newcommand{\caseRule}{{\color{black!14}\rule{\linewidth}{0.4pt}}}
\newcommand{\caseLit}{{\bfseries\color{black!60}Literature}}
\newcommand{\caseOurs}{{\bfseries FlatSeek}}
\node[card] (casea) {%
  {\footnotesize\bfseries (a)\enspace Three parallel boost choppers
  {\mdseries\color{black!55}\cite{messaoudi2023nboost}}}\\[-6pt]
  \caseRule\\[4pt]
  {\scriptsize\(
    \begin{aligned}
      L_i\dot x_i &= V_i-v_c(1-\alpha_i),\quad i=1,2,3,\\
      C\dot v_c &= \textstyle\sum_{i=1}^{3}x_i(1-\alpha_i)-v_c/R.
    \end{aligned}\)}\\[1pt]
  \caseRule\\[3pt]
  {\scriptsize
    \begin{tabular}{@{}l@{\hspace{8pt}}l@{}}
      \caseLit &
        \(y=\bigl(x_1,\ x_2,\ (\textstyle\sum_iL_ix_i^2+Cv_c^2)/2\bigr)^{\!\top}\)\\[1pt]
      \caseOurs &
        \(y=\bigl(x_1,\ Cv_c^2+L_2x_2^2,\ x_3\bigr)^{\!\top}\)
    \end{tabular}}
};

\node[card, below=4pt of casea] (caseb) {%
  {\footnotesize\bfseries (b)\enspace Two-actuator 2D SLIP, stance phase
  {\mdseries\color{black!55}\cite{chen2020slipflat}}}\\[-6pt]
  \caseRule\\[3pt]
  {\scriptsize\(
    \begin{aligned}
      \dot\theta &= \omega,\qquad \dot\ell = v,\\
      \dot\omega &= -2\omega v/\ell-(g\cos\theta)/\ell+u_2/(M\ell^2),\\
      \dot v &= -g\sin\theta+\omega^2\ell+(k_s/M)(\ell_0-\ell+u_1).
    \end{aligned}\)}\\[2pt]
  \caseRule\\[3pt]
  {\scriptsize
    \begin{tabular}{@{}l@{\hspace{8pt}}l@{}}
      \caseLit &
        \(y=\bigl(\ell\cos\theta,\ \ell\sin\theta\bigr)^{\!\top}\)\\[1pt]
      \caseOurs &
        \(y=\bigl(\theta,\ \cos\theta-\ell_0+\ell\bigr)^{\!\top}\)
    \end{tabular}}
};
\end{tikzpicture}
\caption{Two cases from the literature benchmark (SLIP: spring-loaded inverted pendulum). The displayed outputs satisfy Definition~\ref{def:flatness} on their respective regular domains.}
\label{fig:case}
\end{figure}

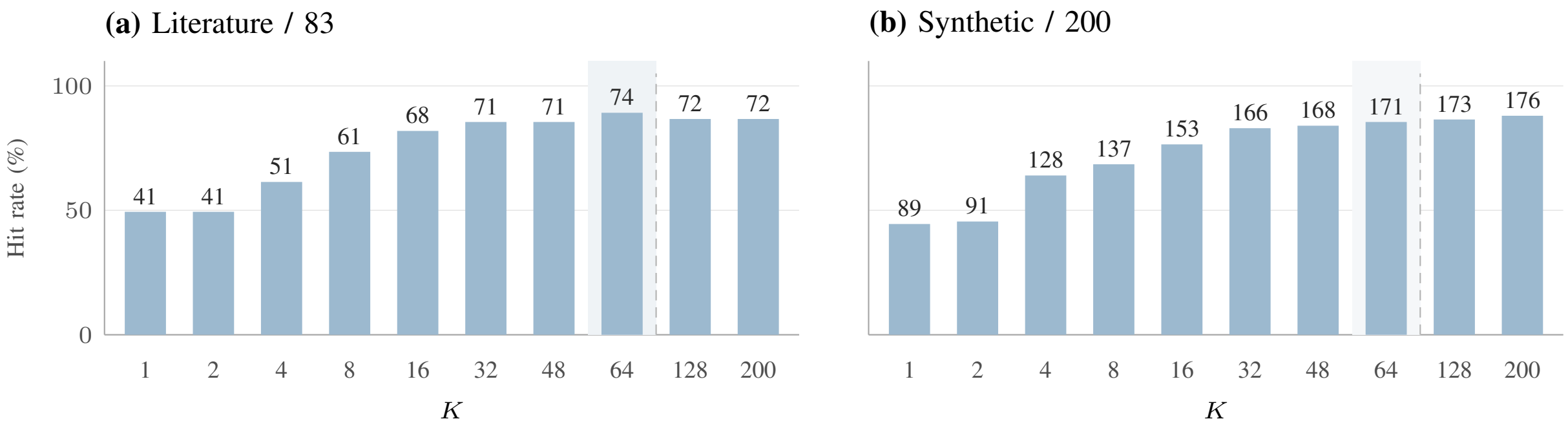
\begin{figure*}[t]
\centering
\begin{tikzpicture}
\definecolor{budgetBlue}{HTML}{37658C}
\definecolor{budgetLightBlue}{HTML}{9CB9CF}
\pgfplotsset{budget panel/.style={
  scale only axis, width=0.40\textwidth, height=2.80cm,
  ybar, bar width=12pt,
  xmin=0.4, xmax=10.6, enlarge x limits=false,
  ymin=0, ymax=110, enlarge y limits=false,
  ytick={0,50,100},
  xlabel={$K$}, xlabel style={font=\scriptsize},
  xtick={1,2,3,4,5,6,7,8,9,10},
  xticklabels={1,2,4,8,16,32,48,64,128,200},
  tick label style={font=\scriptsize, text=black!70},
  axis x line*=bottom, axis y line*=left,
  axis line style={black!35, line width=0.4pt},
  tick style={draw=none},
  axis background/.style={fill=white},
  clip=false,
  ymajorgrids=true, grid style={black!10, line width=0.3pt},
  point meta=explicit symbolic, nodes near coords,
  every node near coord/.append style={
    font=\scriptsize, text=black!85, anchor=south, inner sep=0pt, yshift=2pt},
}}
\begin{axis}[budget panel, at={(0.08\textwidth,0)}, anchor=south west,
  ylabel={Hit rate (\%)}, ylabel style={font=\scriptsize, text=black!70}]
\node[anchor=south west, font=\small, inner sep=0pt]
  at (rel axis cs:0,1.08) {\textbf{(a)} Literature / 83};
\path[fill=budgetBlue!8] (axis cs:7.5,0) rectangle (axis cs:8.5,110);
\addplot[fill=budgetLightBlue, draw=none] coordinates {
  (1,49.4)[41] (2,49.4)[41] (3,61.4)[51] (4,73.5)[61]
  (5,81.9)[68] (6,85.5)[71] (7,85.5)[71] (8,89.2)[74]
  (9,86.7)[72] (10,86.7)[72]};
\draw[black!30, densely dashed, line width=0.4pt]
  (axis cs:8.5,0) -- (axis cs:8.5,105);
\end{axis}
\begin{axis}[budget panel, at={(0.52\textwidth,0)}, anchor=south west,
  yticklabels=\empty]
\node[anchor=south west, font=\small, inner sep=0pt]
  at (rel axis cs:0,1.08) {\textbf{(b)} Synthetic / 200};
\path[fill=budgetBlue!5] (axis cs:7.5,0) rectangle (axis cs:8.5,110);
\addplot[fill=budgetLightBlue, draw=none] coordinates {
  (1,44.5)[89] (2,45.5)[91] (3,64.0)[128] (4,68.5)[137]
  (5,76.5)[153] (6,83.0)[166] (7,84.0)[168] (8,85.5)[171]
  (9,86.5)[173] (10,88.0)[176]};
\draw[black!30, densely dashed, line width=0.4pt]
  (axis cs:8.5,0) -- (axis cs:8.5,105);
\end{axis}
\end{tikzpicture}
\caption{Candidate-budget ablation. Bars show hit rates and labels give hit counts. The shaded bar marks $K=64$, the setting used in Table~\ref{tab:main}.}
\label{fig:abl-k}
\end{figure*}

\subsection{\texorpdfstring{Evaluation Setup}{Evaluation Setup}}\label{sec:exp-protocol}

\textbf{Benchmarks.} The literature benchmark contains $83$ published
systems, and the synthetic benchmark contains $200$ held-out systems generated
by the backward construction in Section~\ref{sec:bg-generation}. Synthetic reconstruction maps certify a flat output, but models
receive only dynamics, not reference outputs.

\textbf{Evaluation protocol.} All language models receive the same prompt (Fig.~\ref{fig:prompt}), use the officially recommended temperature and decoding settings of their base models, and check at most $K=64$ candidates per system. Traditional methods run symbolic search with a limit of $120$\,s per system. Candidates from all methods are judged by the same verifier, and candidates it cannot decide count as failures.

\textbf{Baselines.}
We implement and evaluate five traditional symbolic methods: moving-frame construction~\cite{levine2009flatness,levine2011nsc}; static feedback linearization~\cite{hunt1983global}; saddle--Jacobi screening~\cite{kaminski2024jacobi}; bounded pure prolongation~\cite{charlet1989dynamic,levine2025prolongation}; and state-coordinate enumeration. The saddle--Jacobi baseline enumerates $m$-element subsets of the state and input coordinates and screens them using an order bound and a truncated determinant. Untuned reference models comprise Qwen3.5-2B/9B/27B~\cite{qwen2026qwen35} and R1-Distill-32B~\cite{deepseek2025r1}, together with the API reasoning model DeepSeek V4 Flash~\cite{deepseek2026v4}.

\textbf{Our configuration.} We fine-tune Qwen3.5-2B with LoRA ($r=32,\alpha=64$) for $1500$ steps, using an effective batch size of $16$, a learning rate of $2\times10^{-5}$, and a maximum sequence length of $2048$. Training data consist of backward-generated labeled systems. The base corpus contains $12{,}375$ accepted system--output pairs; after the augmentation in Section~\ref{sec:train-supervision}, the SFT corpus contains $37{,}125$ rows. Reconstruction maps are retained as certificates for the generated labels but are not provided to the model as inputs or targets. The $200$ synthetic test systems are held out from this corpus and are structurally disjoint from it: none shares a structural fingerprint with a training system.

\subsection{\texorpdfstring{Flat-Output Discovery Results}{Flat-Output Discovery Results}}\label{sec:exp-main}

Table~\ref{tab:main} shows that \method{} obtains $74/83$ hits on the literature
benchmark and $171/200$ hits on the synthetic benchmark under the reported candidate-checking protocol. The improvement over untuned Qwen3.5-2B ($50/83$ and $28/200$) reflects the complete trained search configuration. Figure~\ref{fig:case} presents alternative symbolic flat outputs not reported in the literature, generated without supplying the reported outputs. The strongest traditional baselines obtain
$58/83$ and $127/200$ hits on the literature and synthetic benchmarks, respectively.
Untuned hits barely change from 9B to 27B ($59\to61$, $88\to89$).

\subsection{\texorpdfstring{Candidate Budget Ablation}{Candidate Budget Ablation}}\label{sec:exp-ablation}

Figure~\ref{fig:abl-k} varies the candidate budget $K$, each in an
independent sampling run. At $K=1$, \method{} already solves $89/200$
synthetic systems. Hits rise quickly up to $K=16$ (literature
$41\to68$, synthetic $89\to153$) and level off beyond $K=64$, the setting
used in Table~\ref{tab:main}.

\section{Conclusion}\label{sec:conclusion}

We presented \method{} for symbolic flat-output discovery. By
constructing dynamics backward from finite-derivative reconstruction maps, we
obtain labeled systems for task-specific language-model training. The trained
model proposes multiple candidates for verification. On the literature and
synthetic benchmarks, \method{} solves $74/83$ and $171/200$ systems,
compared with $58/83$ and $127/200$ for the strongest traditional method, and it uncovers symbolic flat outputs not reported in the literature (Fig.~\ref{fig:case}). Future work will extend the search to systems for which existing methods have not yet produced a flat-output candidate.

\bibliographystyle{IEEEtran}
\bibliography{references}
\end{document}